# Complex nuclear-structure phenomena revealed from the nuclide production in fragmentation reactions

*M. V. Ricciardi[1,§], A. V. Ignatyuk[2], A. Kelić[1], P. Napolitani[1,3],
F. Rejmund[3], K.-H. Schmidt[1], O. Yordanov[1]*

[1]GSI, Planckstr. 1, 64291 Darmstadt, Germany
[2]IPPE, Bondarenko Squ. 1, 249020 Obninsk, Russia
[3]IPN, 91406 Orsay, France

**Abstract:** Complex structural effects in the nuclide production from the projectile fragmentation of 1 $A$ GeV $^{238}$U nuclei in a titanium target are reported. The structure seems to be insensitive to the excitation energy induced in the reaction. This is in contrast to the prominent structural features found in nuclear fission and in transfer reactions, which gradually disappear with increasing excitation energy. Using the statistical model of nuclear reactions, relations to structural effects in nuclear binding and in the nuclear level density are demonstrated.

## 1. Introduction

Nuclear structure manifests itself in many features, which are widely investigated, e.g. in ground-state properties like binding energy, half-life, radius and shape. Systematic measurements of the total nuclear reaction cross section [1] have been an important tool in the study of halo nuclei. Signatures of nuclear structure arise also in the production yields in specific nuclear reactions at low energies. The enhanced production of even elements [2, 3] and the appearance of fission channels in low-energy fission [4, 5], as well as the structural features observed in transfer reactions [6] are typical examples of these signatures. These structures gradually disappear and transform into smooth distributions with increasing excitation energy induced in the reaction. The energy dependence of even-odd effects [7] and shell structure [8, 9] in fission has been understood in the framework of the statistical model.

In the last years, signatures of nuclear structure were found in the production yields in deep-inelastic and in fragmentation reactions, which can be quite violent and which are expected to introduce a large amount of excitation energy in the nucleus. This experimental result can have a two-fold interpretation: either nuclear structure can manifest itself also in the end-products of very hot nuclei, or part of the reaction, by some unknown reason, passes by very low excitation energies.

In the present work, we report on the production yields from the projectile fragmentation of 1 $A$ GeV $^{238}$U nuclei in a titanium target, measured at GSI. The residual nuclei were fully identified in mass and atomic number with the high-resolution magnetic spectrometer FRS, and their production cross sections were deduced. This is the first time that this kind of fine structure in the nuclide production is systematically investigated with full nuclide identification over an extended area of the chart of the nuclides. We will discuss the reason for this structure and offer our interpretation in the frame of the statistical model.

---

[§] This work forms part of the PhD thesis of M. V. Ricciardi.

## 2. Experimental production cross sections

In several experiments, in which different rather violent nuclear reactions have been investigated, a fine structure in the nuclide production, manifested as an even-odd effect, has been observed, see e.g. [10, 11, 12, 13, 14, 15, 16, 17, 18]. Most experiments could determine the nuclear charge of the reaction products, only. Consequently, only the enhancement in the production of even-$Z$ elements, found in the order of a few tens per cent, could be investigated. Lately, with the use of spectrometers, also the neutron number became accessible: the most remarkable finding of these more recent experiments was a variation of the magnitude of this fine structure with the mean neutron excess of the reaction products [13, 18].

In the present work, we report on new results [19] obtained in the reaction $^{238}$U + Ti at 1 $A$ GeV in an experiment at the fragment separator (FRS) at GSI. The FRS [20] is a two-stage, high-resolution, magnetic forward spectrometer. It has a dispersive intermediate image plane and an achromatic image plane at the exit. The 1 $A$ GeV $^{238}$U primary beam impinged on a 36 mg/cm$^2$ titanium target, placed at the entrance of the FRS. The primary-beam intensity of about $10^7$ particles per second was constantly monitored. Two scintillation detectors, placed at the intermediate plane and at the final plane (see Figure 1), were used to detect the horizontal position and the time-of-flight. The two horizontal positions gave a measurement of the radius, $\rho$, of the fragment trajectory, and, with the measurement of the magnetic fields inside the dipoles, $B$, the magnetic rigidity, $B\rho$, was obtained. Along with the measured velocity, obtained from time-of-flight and flight path, the $A/Z$-ratio of the fragment was determined from the equation:

$$\frac{A}{Z} = \frac{e}{m_0} \cdot \frac{B\rho}{\gamma \beta c} \qquad (1)$$

where $\beta c = \upsilon$ is the velocity of the ion, $\gamma$ is the relativistic parameter, $m_0$ is the nuclear mass unit, and $-e$ is the electron charge. Two ionisation chambers, placed behind the FRS, recorded the energy loss of the produced ions, giving a measurement of the nuclear charge, $Z$. Multi-wire proportional counters (MWPCs) provided additional tracking information. The experimental procedure and the performed data analysis correspond to those used already in previous experiments, and more details can be found in Ref. [21].

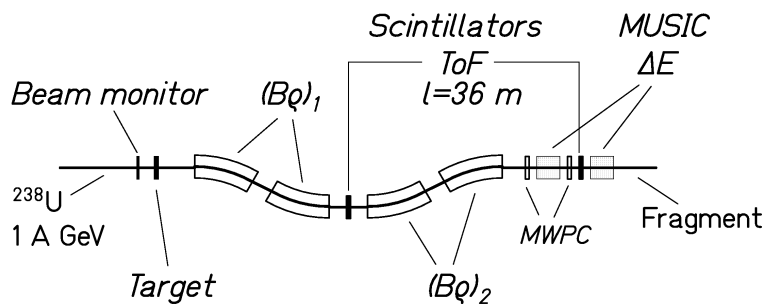

**Figure 1:** Schematic drawing of the FRS with the detectors used in the experiment.

The good resolution obtained in the fragment identification can be observed in Figure 2, where the characteristic pattern $Z$ vs. $A/Z$ is presented. The sequence along the vertical line at $A/Z=2$, which collects the yields of fragments with $N=Z$, shows an enhanced production of even-$Z$ nuclei. Looking carefully at the yields of nuclei with $N-Z=5$, it can be observed that in this case the odd-$Z$ nuclei show an enhanced production. This observation suggested us to filter the data according to the neutron excess $N-Z$. The production cross sections of the observed fragments, grouped according to this filter, are shown in Figure 3. The data reveal a complex structure. All even-mass nuclei present a visible even-odd effect, which is particularly strong for $N=Z$ nuclei. Odd-mass nuclei show a reversed even-odd effect with enhanced production of odd-$Z$ nuclei. This enhancement is stronger for nuclei with larger values of $N-Z$. However, for nuclei with $N-Z=1$ the reversed even-odd effect vanishes out at about $Z=16$, and an enhanced production of even-$Z$ nuclei can again be observed for $Z > 16$. Finally, all the observed structural effects seem to vanish out as the mass of the fragment increases.

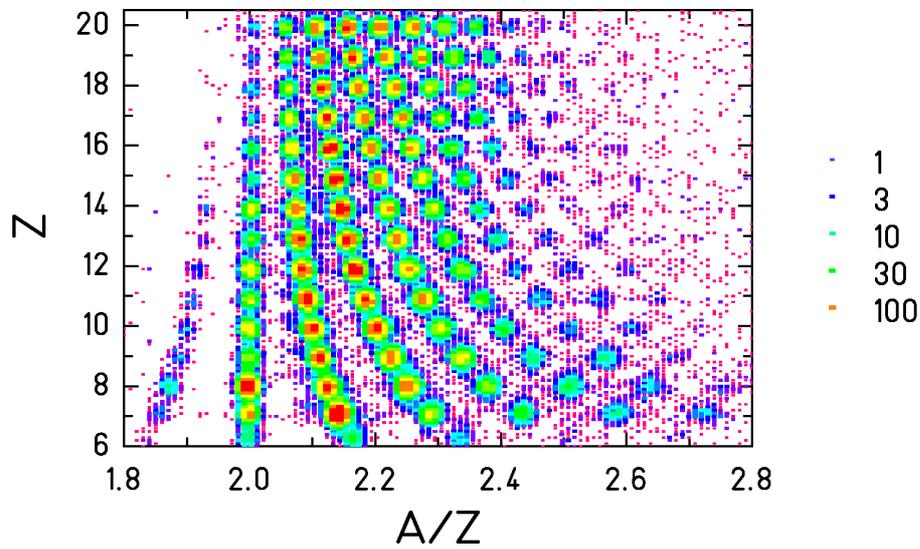

**Figure 2:** Cluster plot of $Z$ versus $A/Z$ of the projectile-like reaction products from the reaction $^{238}$U + Ti at 1 $A$ GeV.

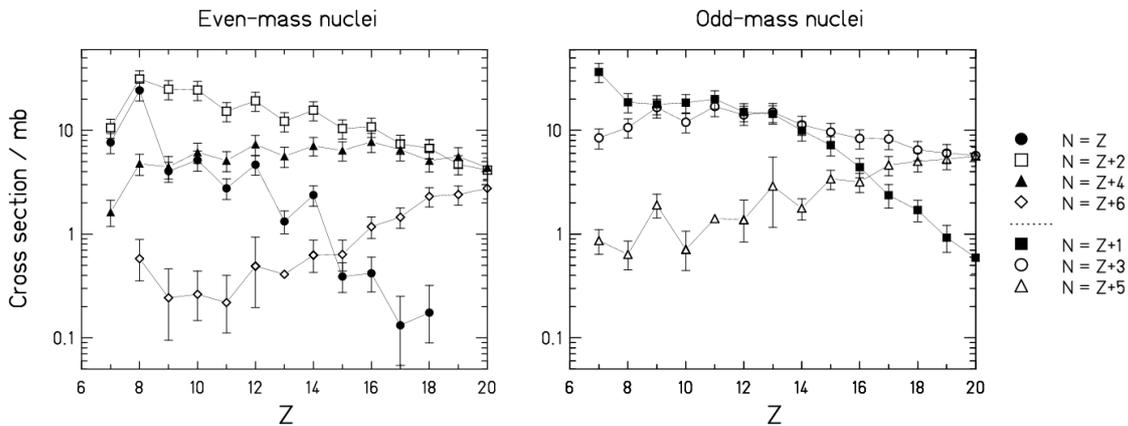

**Figure 3:** Formation cross sections of the projectile-like residues from the reaction $^{238}$U + Ti, 1 $A$ GeV. The data are given for specific values of $N-Z$. The cross section for $^{32}$Al ($Z=13$, $N=Z+6$) is an extrapolated value. The chain $N=Z$ shows the strongest even-odd effect, while the chain $N-Z=5$ shows the strongest reversed even-odd effect.

This complex behaviour is summarised in Figure 4, where the local relative even-odd effect, $\delta_{rel}$, is quantitatively evaluated by means of the equation:

$$\delta_{rel}(Z+3/2) = \frac{1}{8}(-1)^{Z+1}[\ln Y(Z+3) - \ln Y(Z) - 3(\ln Y(Z+2) - \ln Y(Z+1))] \qquad (2)$$

where $Y(Z)$ is the value of the production cross section for the nucleus with charge $Z$ and with given $N-Z$ value. The above equation is the standard prescription proposed by Tracy [22], which describes the local deviation of the cross sections from a Gaussian-like distribution. $\delta_{rel}(Z)$ is a quantity measured over four consecutive cross sections centred at $(Z+3/2)$. A value equal to zero means a smooth behaviour, a positive value means enhanced production of even-$Z$ nuclei, a negative value means enhanced production of odd-$Z$ nuclei. In the range covered by the data, the sequence with $N=Z$ shows the strongest effect, reaching values of the order of 50%. Thus, this structure is even stronger than any even-odd structure observed in low-energy fission [2]. Other even-mass nuclei show a much weaker effect, hardly exceeding 10%. For odd-mass nuclei, the intensity of the reversed even-odd effect is strongest for $N-Z=5$ nuclei. It decreases gradually with decreasing $N-Z$ number and changes sign for the heavier nuclei with $N-Z=1$.

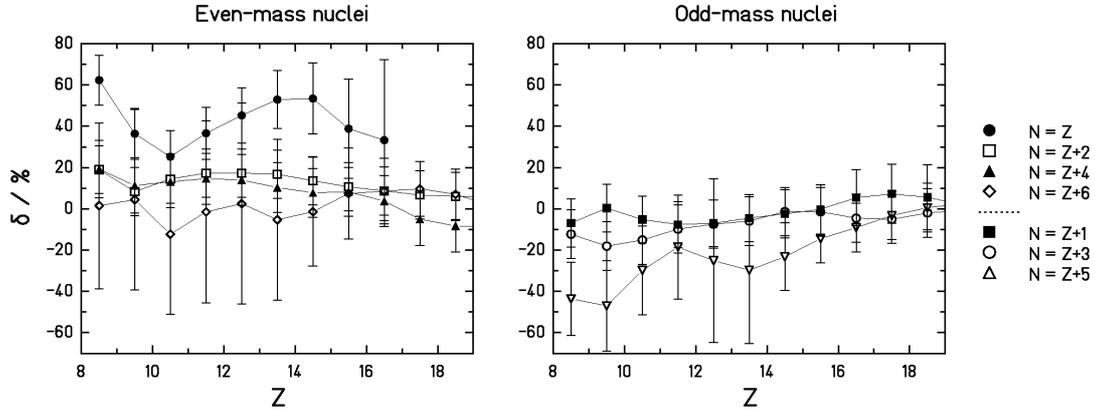

**Figure 4:** Local relative even-odd effect in the cross sections reported in Figure 3, calculated with the use of Equation (2).

Summing up all the cross sections for the isotopes of every element, Figure 5 could be constructed. There, the even-odd structure is still visible, although less pronounced. Thus the present experiment is in agreement with the results reported in Refs. [10, 11, 12, 13, 14, 15, 16, 17, 18], but it provides more detailed information.

## 3. Analysis with a simple statistical model

In the production of light nuclides, even-odd fluctuations of similar magnitude of the ones found in the present work have been observed in a variety of different reactions for many different projectile-target combinations and a wide range of beam energy. Some examples are listed in Table 1. In a recent study, M. Balasubramaniam *et al.* were able to reproduce structural effects in the formation cross sections of intermediate mass fragments with the dynamical cluster-decay model [23]. However, the apparent insensibility of the even-odd staggering in the production cross sections of light residues to the type of nuclear reaction suggested us to assume that these even-odd fluctuations are produced in the later stage of the deexcitation process.

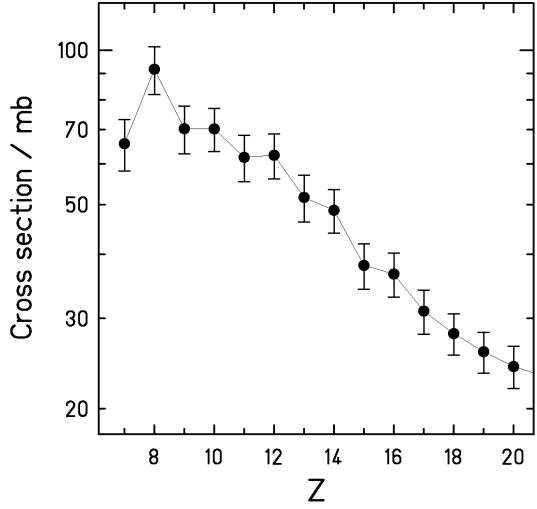

**Figure 5:** Nuclear-charge distribution for the lightest uranium fragmentation residues observed in the reaction $^{238}$U + Ti at 1 $A$ GeV. The distribution shows an even-odd structure of the produced elements.

**Table 1:** Previous experiments with signatures of fine structure after the de-excitation of highly excited nuclei.

| Reference | Reaction | Beam energy [$A$ MeV] |
|---|---|---|
| Sl. Cavallaro et al. [16] | $^{35}$Cl + $^{24}$Mg | 8 |
| E. M. Winchester et al. [18] | $^{40}$Ca + $^{58}$Ni <br> $^{40}$Ar + $^{58}$Fe | 25 |
| Ch. O. Bacri et al. [13] | $^{40}$Ar + Ni | 44 |
| L. B. Yang et al. [17] | $^{58}$Fe + $^{58}$Fe <br> $^{58}$Ni + $^{58}$Ni | 45 to 105 |
| B. Blank et al. [11] | $^{40}$Ar + $^{12}$C | 403 |
| C. N. Knott et al. [14] | e.g. $^{32}$Si + $^{1}$H | e.g. 571 |
| W. R. Webber et al. [12] | $^{56}$Fe + $^{12}$C | 600 |
| C. Zeitlin et al. [15] | $^{56}$Fe + div. | 1050 |
| A. M. Poskanzer et al. [10] | $^{238}$U + $^{1}$H | 5500 |

In this Section, we want to test the hypothesis that the even-odd fluctuations are produced at the end of the evaporation cascade due to the influence of nuclear structure on the number of bound states. To do this, we make use of a simple statistical model, with the manifestation of pairing taken into account in a schematic way, both in the binding energies and in the level densities. The masses are calculated with the liquid-drop model of Myers and Swiatecki [24] without shell and pairing terms. The effect of pairing was reinserted by modulating the binding energies by an even-odd staggering quantified by $\Delta \approx 12/\sqrt{A}$ MeV, in such a way that nuclei with an even proton (or neutron) number are on average more bound by $\Delta$ than nuclei with an odd proton (or neutron) number. The density of excited states, $\rho$, is calculated with the well-known Fermi-gas formula [25]:

$$\rho(E) = \frac{\sqrt{\pi}}{12} \frac{\exp(2\sqrt{a(E-\delta)})}{a^{1/4}(E-\delta)^{5/4}} \tag{3}$$

where $a$ is the level-density parameter taken equal to $A/9$ MeV$^{-1}$, $E$ is the excitation energy of the nucleus and $\delta$ is the even-odd pairing correction equal to 0, $\Delta$ or $2\Delta$ for odd-odd, odd-even or even-odd, and even-even nuclei, respectively. The number of final states is determined by the number of states available between the ground state and the threshold, which is given by the minimum value of the neutron separation energy, $S_n$, and the proton separation energy, $S_p$. The Coulomb barrier for proton emission was neglected for these light nuclei.

The number of final states, obtained with this method, is shown in Figure 6. The analysis is aimed just to compare the local relative even-odd structure. The simple model that we used can only indicate whether structural effects can be restored at the end of the decay chain, in the last step of the de-excitation process, which is dominated by the available phase space of the final residue. The number of possible final states must not be compared with the global tendencies of the cross sections since other effects, connected to the reaction mechanisms, affect the final result and have to be taken into account for a quantitative description.

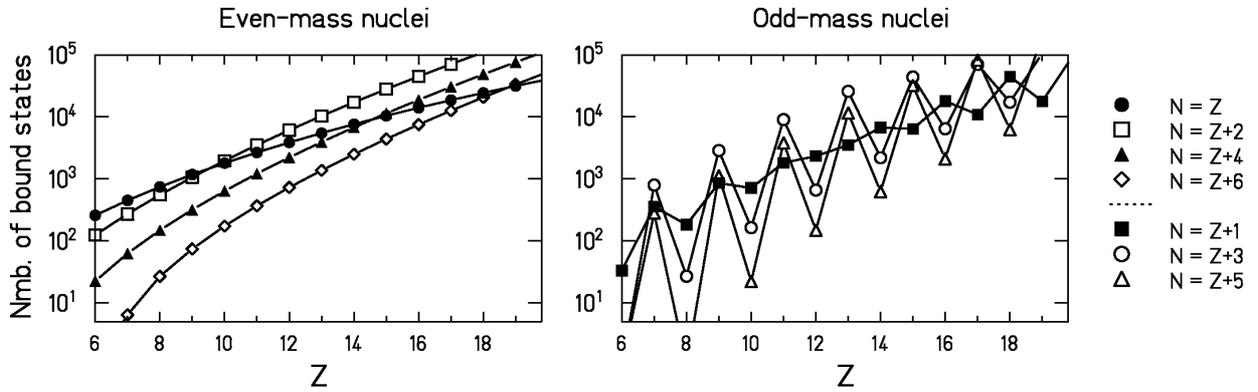

**Figure 6:** Number of bound states, representing the number of possible final states, determined by the number of energy levels available between the ground state and the lowest particle-decay threshold.

For the odd-mass nuclei, the statistical model reproduces the observed structural effects in all their complexity. In particular, the reversed even-odd effect evaluated along constant $N$-$Z$ lines is reproduced, and it even increases with increasing neutron excess. With increasing the nuclear charge, the strength of the reversed even-odd effect decreases and after a certain $Z_T$ value even-$Z$ nuclei become more produced than odd-$Z$ nuclei. The $Z_T$ value increases as $N$-$Z$ increases. For $N$-$Z$=1 nuclei, the model predicts $Z_T$=13, a value rather close to the one found in the experimental data. Finally, the structural effects vanish out as the mass of the fragment increases.

The agreement between the predictions of this simple model and the measured data for odd-mass nuclei is easily understandable with the help of Figure 7, which shows the experimental ground-state energies $E_{gs}$ [26], and the absolute particle separation energies[1] with respect to the liquid-drop ground-state energy [24]. The ground-state energies calculated with the liquid-drop model do not contain any even-odd nor shell effect. The structures appearing in $E_{gs}$-$E_{ld}$ (full dots in Figure 7) result from the shell effects in the experimental ground-state masses.

---

[1] As absolute particle separation energy we take the sum of the experimental ground-state energy $E_{gs}$ and the experimental particle separation energy $S_v$, both taken from Ref. [26].

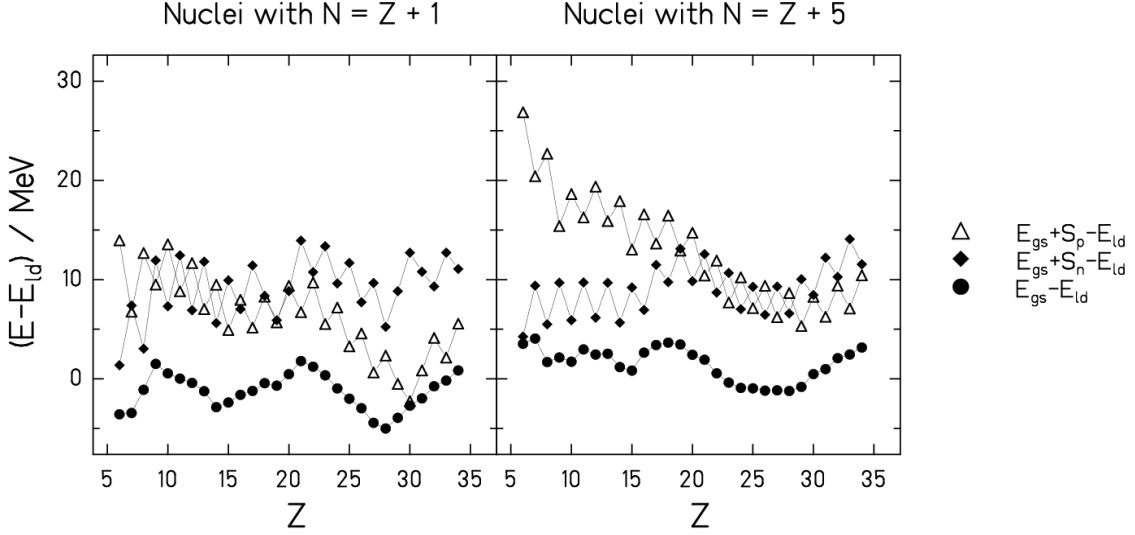

**Figure 7**: The energy of the ground state $E_{gs}$ and the absolute separation energy defined as $E_{gs}+S_v$, for $N=Z+1$ and $N=Z+5$ nuclei, above the energy of the corresponding liquid-drop [24] ground state $E_{ld}$. The compilation of the data is taken from Ref. [26]. The symbols are defined in the figure.

First of all, Figure 7 clearly shows that the staggering effect in the number of bound states in the chains of odd-mass nuclei results from the staggering of the separation energies above the smooth chain of ground states. Secondly, by comparing chains with $N=Z+1$ and $N=Z+5$ we see that the values of $S_p$ systematically increase with increasing neutron excess and decrease with increasing mass, while the opposite trends are observed for the $S_n$ values. Figure 7 shows that moving along a given $N-Z$=odd chain, the lowest particle separation energy passes from neutron separation energy to proton separation energy as the nuclear charge of the fragment increases, which explains the transition from the a reversed even-odd effect with enhanced production of odd-Z and even-N nuclei to a positive even-odd effect with enhanced production of even-Z and odd-N nuclei. The transition point moves to higher $Z$ values with increasing value of $N-Z$.

This good reproduction of the qualitative features of the staggering of the experimental data for odd-mass residues encourages us to believe that the structural effects are restored in the end-products of hot decaying nuclei, and that the structure is ruled by the available phase space at the end of the de-excitation process. For the even-mass nuclei, however, this approach does not explain the observed structure. Therefore, we will refine our analysis in the next Section.

## 4. Analysis with an evaporation model

We have seen in the previous Section that the number of bound states in the final nucleus can only partly explain the observed even-odd staggering in the production cross sections. In this Section we want to check if the last stages of the evaporation cascade could have an influence on the observed structure.

### 4.1. The abrasion-ablation model

To obtain a more complete description of the observed even-odd effects, we have calculated the yields of light nuclei produced in the reaction of $^{238}$U with Ti at 1 $A$ GeV with a statistical abrasion-evaporation model, realised within the ABRABLA code [27, 28, 29]. In ABRABLA, after the nucleus-nucleus interaction, described as an abrasion process, the pre-fragment at every step of its evolution has two possible decay channels: evaporation and fission.

The probability that a compound nucleus ($Z,N$) with excitation energy $E$, emits the particle $\nu$ is given by:

$$P_\nu(E) = \frac{\Gamma_\nu(Z,N,E)}{\sum_i \Gamma_i(Z,N,E)} \qquad (4)$$

where $i$ denotes all the possible decay channels (specifically: neutron emission, proton emission, alpha emission, fission). The particle emission width $\Gamma_\nu$ can be written as [30]:

$$\Gamma_\nu(E) = \frac{1}{2\pi\rho_c(E)} \frac{4m_\nu R^2}{\hbar^2} T^2 \rho_d(E - S_\nu - B_\nu) \qquad (5)$$

where $m_\nu$ denotes the particle mass, $S_\nu$ is the separation energy, $B_\nu$ is the effective Coulomb barrier, $R$ is the radius of the nucleus, $T$ is the temperature of the residual nucleus after particle emission, $\rho_c$ and $\rho_d$ are the level densities of the compound nucleus and the daughter nucleus, respectively. The fission widths are calculated in a similar way taking into consideration the transient-time effects with an approach described in Ref. [31].

To be consistent in the description of pairing effects in the masses and the level densities, we have initially performed the calculations of the masses and separation energies with the liquid-drop model of Myers and Swiatecki [24] and the pairing gap equal to $\Delta \approx 12/\sqrt{A}$ for both the masses and the level densities. The results obtained are shown in Figure 8.

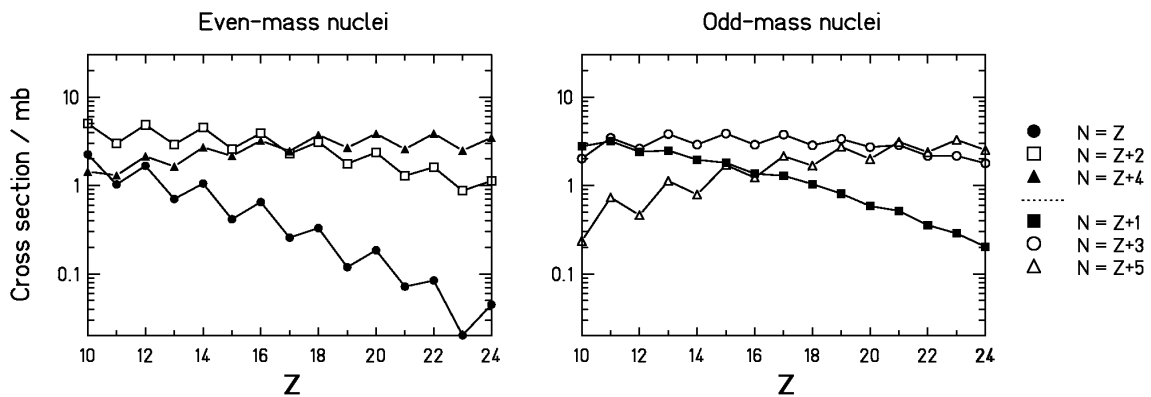

**Figure 8:** Production cross sections of fragmentation residues from the reaction $^{238}$U +Ti at 1 $A$ GeV, calculated with the statistical code ABRABLA taking into account the pairing effects in binding energies and level densities.

The most interesting result is the fact that the even-odd effect for the $N$-$Z$=even chains is quite well reproduced by the statistical abrasion-ablation code, contrary to results of the calculation with the simple statistical model described in Section 3.

The main difference between ABRABLA and the simple statistical model is that in the full calculation at each step the probability of a certain decay channel is not only determined by the number of possible final bound states, but also by the number of possible excited levels in which the mother nucleus can reside before entering the next decay channel. It is exactly for this reason that structural effects due to pairing are restored for even-mass nuclei. This can be better understood with the help of Figure 9.

Let us consider two neighbouring odd-mass mother nuclei, an odd-even ($Z$=odd, $N$=even) and an even-odd ($Z$=even, $N$=odd) nucleus, that decay by e.g. neutron evaporation. In both nuclei, the distributions of level densities are quite similar, see Equation 3. By evaporating a neutron, an odd-even mother nucleus decays into an odd-odd daughter (left part of Figure 9), and an even-odd mother decays into an even-even daughter (right part of Figure 9). The ground states of the daughter nuclei, compared to the ground states of the corresponding mother nucleus, are shifted by the neutron separation energy of the respective mother nucleus $S_n^{mother}$. Moreover, because of the pairing interaction, the ground state of the even-even daughter is lowered by $2\Delta$ compared to the ground state of the odd-odd daughter. For any excited level above $S_n^{mother}$, the mother nucleus can decay into a level of the daughter nucleus. According to the simple picture from Section 3, the probability to create a given daughter nucleus is determined only by the number of available levels between the ground state and the neutron separation energy of the daughter nucleus $S_n^{daughter}$. Although the ground state of the even-even daughter is lower than the ground state of the odd-odd daughter, the number of bound states in these two nuclei is practically the same. As a consequence, according to the simple statistical model, there is no difference in the population of an odd-odd and an even-even daughter nucleus.

However, in the full description of the evaporation process, the probability to create a given daughter nucleus is determined not only by the number of the bound levels in the daughter nucleus but also by the number of levels occupied by the mother nucleus that are above $S_n^{mother}$. From Figure 9 we can see that the levels in an even-odd mother nucleus decaying into particle-bound levels of the even-even daughter nucleus extend over an energy range that is larger by $2\Delta$ than in the case of an odd-even mother nucleus. Therefore, although the number of bound levels in the even-even and the odd-odd daughter are almost the same, the probability that a particle-bound level in the even-even daughter is populated is higher than in the odd-odd daughter because there are more available levels in the even-odd mother than in the odd-even mother nucleus. As a consequence, according to the full evaporation model the experimentally observed even-odd staggering is also reproduced for even-mass nuclei (chains with even $N-Z$ values).

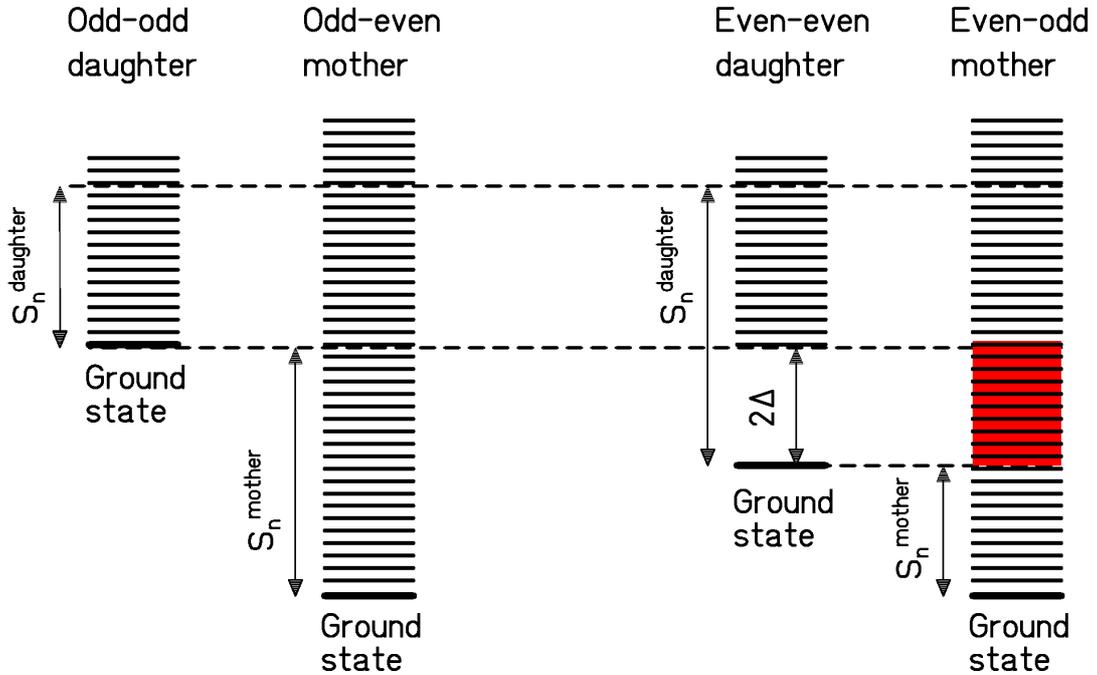

**Figure 9:** Schematic pattern (not to scale) of the levels for two different combinations of mother-daughter nuclei decaying by neutron evaporation. When the even-odd mother nucleus occupies one of the levels indicated by the shadowed area, it can decay into the ground state of the even-even daughter nucleus.

In the above reasoning, the kinetic energy of the emitted neutron is neglected, and therefore the limits defining the particle-bound levels ($S_n^{daughter}$ and $S_n^{mother}$) in the daughter nucleus appear as sharp limits (Figure 9). Because the emitted particle has some final velocity, these limits have a certain width, which is correlated with the temperature of the system. As we have considered neighbouring nuclei at the last stages of the evaporation process, these widths should be very similar in both mother nuclei and not large. Therefore, the general conclusion from the above considerations will not be changed.

To consider in what way the even-odd staggering can be modified by the shell effects, we performed the ABRABLA calculations with shell corrections included in both the binding energies and the level densities. The shell corrections were calculated as prescribed by Myers and Swiatecki [32], and the level densities were calculated as described in Ref. [28]. The results obtained are shown in Figure 10. From the comparison with Figure 8 we can conclude that the shell effects do not influence the even-odd staggering, but they change essentially the ratios of the production cross sections of nuclei with different values of *N-Z*.

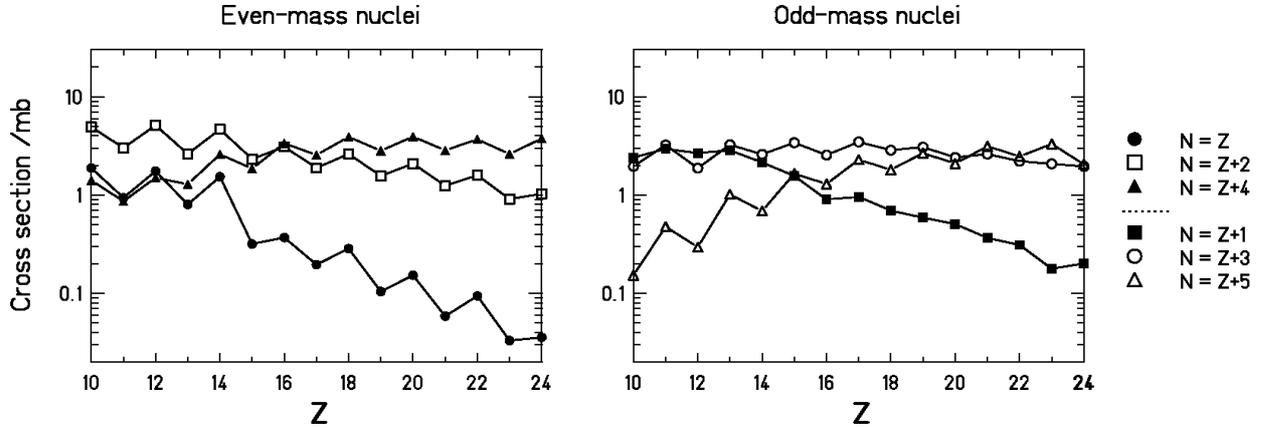

**Figure 10:** Production cross sections of fragmentation residues from the reaction $^{238}$U +Ti at 1 *A* GeV, calculated with the ABRABLA code taking into account pairing effects and shell corrections in the binding energies and the level densities.

Although the full evaporation calculations reproduce the even-odd staggering not only in the chains of odd-mass nuclei but also in the chains of even-mass nuclei, there are still two questions that remain open. The first considers the vanishing of the staggering effects when the fragment mass increases, which is seen in the experimental data but not in the calculations. The second question is that according to the calculations the strength of the even-odd effect in the even-mass nuclei is constant for all the chains, while in the experimental data the *N=Z* chain shows a particularly strong staggering. We will discuss these two features in Subsection 4.2 and in Section 5, respectively.

**4.2. The disappearance of the even-odd effects for heavy residues**

The experimental results presented in Figures 4 and 5 show that the staggering effect in the cross sections rapidly vanishes away as the mass of the fragment increases. Furthermore, there are a lot of experimental results in literature showing that the yields of heavy fragmentation and spallation residues present a smooth behaviour (e.g. [21, 33, 34, 35]).

There are, in our opinion, two main reasons for this. The first one is that the pairing gap $\Delta$ decreases with increasing mass of the nucleus: $\Delta \approx 12/\sqrt{A}$ MeV. The second reason is that gamma emission becomes competitive to particle decay for heavy compound nuclei. This can be better understood again with the help of Figure 9. When the even-odd mother nucleus occupies one of the levels in the shadowed region, it can decay into the ground state of the even-even daughter nucleus. As we have discussed above, this is the reason for the enhanced production of even-even nuclei with respect to the odd-odd ones. However, from the considered energy range of the mother, there is only one possible final state available for the particle-decay channel: the ground state of the daughter nucleus. Nevertheless, the even-odd mother nucleus can also de-excite by emitting a gamma and fall into a lower energy level. In general, the emission of a gamma is appreciably less probable than the particle decay, if the numbers of final states are comparable. For a heavy nucleus, the number of final states available after gamma emission from a state near the particle threshold (the levels between the mother ground-state and $S_\nu^{mother}$) can be very large. In that case, gamma emission and particle decay into the ground state of the daughter nucleus can be two competitive channels. If the number of final states available after gamma emission is very large, then the γ-radiation rate dominates, and the mother nucleus survives, washing out the enhancement of the production of even-even daughter nuclei. Due to the mass dependence of the level density, the number of levels available after the gamma emission increases rapidly with increasing mass of the fragment.

In order to verify the effects of γ-radiation on the final cross-section distribution, the γ-decay channel was included in the ABRABLA code. As the emission of statistical γ-rays occurs predominantly via the giant dipole resonance, its rate can be written in the following way [36]:

$$\Gamma_\gamma(E) = \sum_{I=|J-1|}^{J+1} \int_0^E \varepsilon_\gamma^3 \cdot k(\varepsilon_\gamma) \cdot \frac{\rho(E-\varepsilon_\gamma, I)}{\rho(E, I)} d\varepsilon_\gamma \qquad (6)$$

where $E$ is the excitation energy of the mother nucleus and $k(\varepsilon_\gamma)$ is the radiative strength function for a dipole electric transition. As already said, for high excitation energy the γ emission is negligible compared to the particle emission, and it becomes important only at the energies around and below the particle separation energies. Taking $E=S_n$, and using the power approximations for the radiative strength function [37] and the constant-temperature model [38], Equation (6) can be parameterised as [36]:

$$\Gamma_\gamma(S_n) = 0.624 \cdot 10^{-9} \cdot A^{1.60} \cdot T^5 \quad \text{MeV, with } T = \frac{17.6}{A^{0.699}} \quad \text{MeV} \qquad (7)$$

In the above equation, $A$ is the mass of a mother nucleus and $T$ is the nuclear-temperature parameter of the constant-temperature model [36].

The results of these calculations are shown in Figure 11 for the case of the reaction $^{208}$Pb + $^1$H at 1 $A$ GeV. The figure compares measured isotopic cross sections of $_{71}$Lu from Ref. [21] with two sets of calculations performed with and without taking γ emission into account. Without considering the emission of γ-rays, the partial cross sections show a strong even-odd staggering (open squares) that is not visible in the experimental data. On the other hand, after including γ emission, this staggering is strongly reduced, and the calculations are in much better agreement with the data. The same is found for any other heavy fragment residue, $_{71}$Lu is just one example. This comparison also shows that for the smoothening of the even-odd staggering in the production cross sections, γ emission has a stronger influence than the decrease of the pairing gap with increasing mass.

## 5. A closer view on binding energies and excited levels

In the previous sections, the most salient features of the observed structural effects in the production yields of fragmentation residues were reproduced by using well established but rather schematic descriptions of pairing in binding energies and level densities. Pairing was included by increasing the binding energies and shifting the energy levels by 0, Δ or 2Δ. In this Section we discuss indications from binding energies and spectroscopy for phenomena which go beyond this simple description.

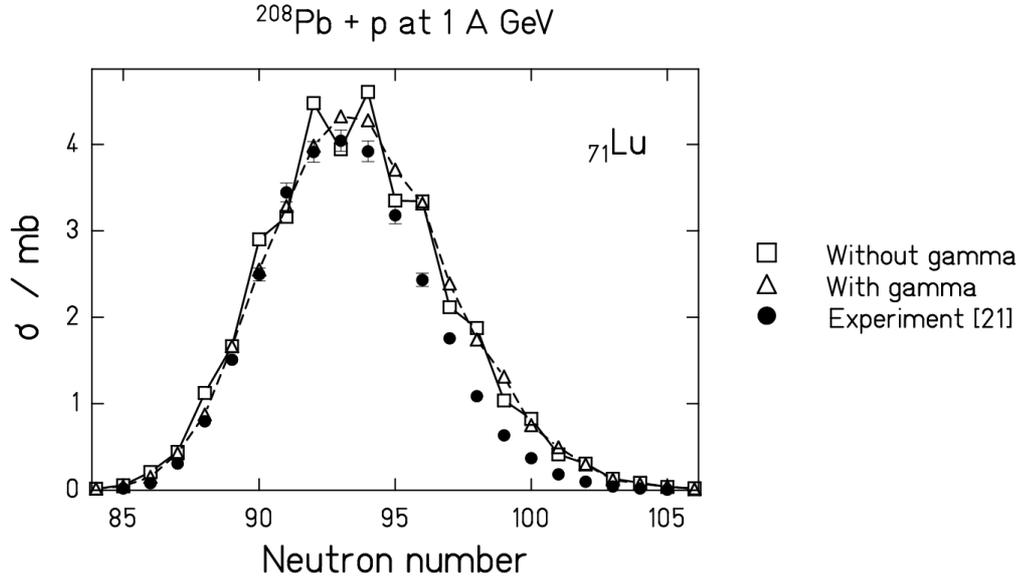

**Figure 11**: Production cross sections of different isotopes of $_{71}$Lu in the reaction $^{208}$Pb (1 $A$ GeV) + $^{1}$H. The experimental data from Ref. [21] are shown as full dots and compared with two sets of calculation: without including γ emission (open squares) and with including it (open triangles). The errors on the experimental data are shown only if the error bars are larger than the symbol size.

**5.1. Binding energies**

It is known that the nuclear binding energies are modulated by an even-odd staggering defined by the pairing gap $\Delta \approx 12/\sqrt{A}$. However, this is only an average value. A detailed analysis of the local absolute even-odd fluctuations in the binding energies of light nuclei is shown in Figure 12 using the following description:

$$\delta_{abs}(Z+3/2) = \frac{1}{8}(-1)^{Z+1}[B(Z+3) - B(Z) - 3(B(Z+2) - B(Z+1))] \quad (8)$$

where $B(Z)$ is the value of the experimental binding energy, taken from reference [26], for the nucleus with nuclear charge $Z$ and with given $N$-$Z$ number. Most nuclei behave as expected. Chains along odd-mass nuclei, which are either even-odd or odd-even, hardly show any even-odd structure. In even-mass nuclei, except those with $N$=$Z$, the different binding energies of even-even and odd-odd nuclei lead to an even-odd effect of around 2 MeV, which slightly decreases with mass, and is only slightly smaller than the global parameterisation $\Delta \approx 12/\sqrt{A}$ MeV. As a remarkable exception, the $N$=$Z$ chain shows a considerably enhanced even-odd structure. This special behaviour of $N$=$Z$ nuclei is caused by the Wigner term [39, 40] in nuclear binding, which is responsible for an additional deficit in the binding energy for $N$=$Z$ odd-odd nuclei of about $30/A$ MeV compared to $N$=$Z$ even-even nuclei [41, 42].

At the same time, the even-odd structure in the production yields of $N$=$Z$ nuclei is also exceptionally strong. Thus, we find a correspondence in the behaviour of the $N$=$Z$ chain, if we compare the even-odd structure in the production cross sections (Figure 4) to the even-odd staggering in the binding energies (Figure 12). However, for a quantitative understanding, also the level densities should be considered.

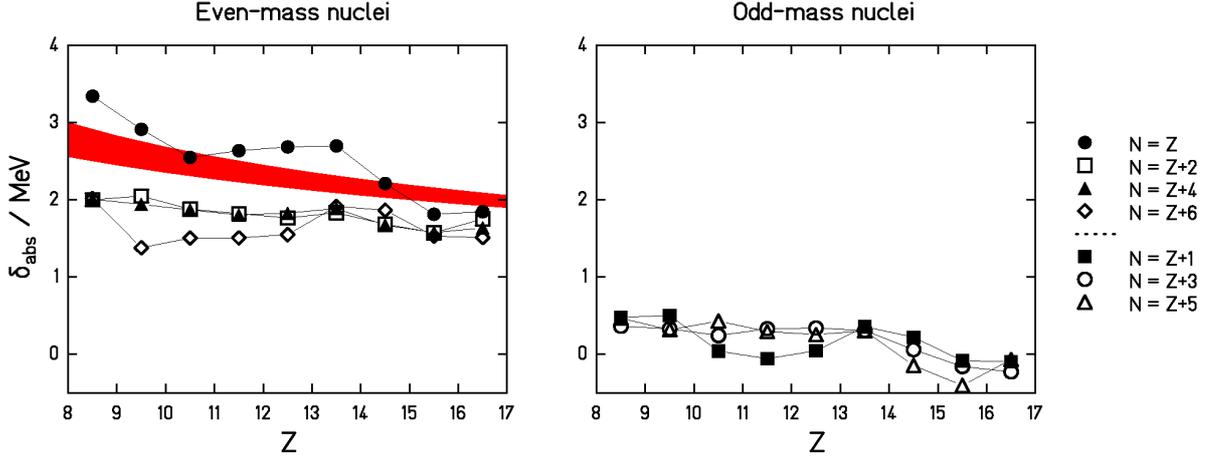

**Figure 12:** Local absolute even-odd effect in the experimental binding energies of Ref. [26], calculated with Equation (8). The values are given along specific cuts in *N-Z*. The values of the pairing gap Δ, calculated as $12/\sqrt{A}$ MeV, fall inside the grey band.

**5.2. Level density**

In this Subsection, we will investigate the influence of pairing correlations on the level density. A good description is offered by the analytic formula of the density $\rho_n$ of *n*-quasi-particle states as a function of excitation energy *E*, which was derived by Strutinsky [43] in the Boltzmann-gas model. For the simplified case of one kind of nucleons (e.g. protons), he obtained:

$$\rho_n = \frac{g^n (E - n\Delta)^{n-1}}{[(n/2)!]^2 (n-1)!} \tag{9}$$

This formula gives a rather accurate formulation of the influence of blocking (the reduction of states involved in pairing correlations due to unpaired particles) on the energies of quasi-particle excitations at low excitation energies. Figure 13 shows the resulting number of levels below a given energy *E-δ* for two nuclei with an even and an odd number of nucleons with the same single-particle level density *g*. The figure reveals an important feature, which is also found in the more complete formulation of the super-fluid-nucleus model (e.g. [44]): When the excitation energy is corrected for the even-odd staggering of the ground-state binding energy, the level densities are almost identical. The shifted level-density formula (Equation (3)) that we used in the previous sections is in agreement with this basic feature of the super-fluid model.

However, Figure 12 clearly demonstrates that the schematic assumption of an even-odd staggering of the binding energies by a pairing gap Δ, which only depends on mass, is not realistic. The influence of nuclear structure on the binding energies is more complex, e. g. the even-odd staggering along the cut *N=Z* is about 1.5 times larger than the staggering along other cuts of even-*A* nuclei. Therefore, it seems to be natural to expect a more complex influence of pairing correlations on the level density than depicted in Figure 13, too.

Figure 14 shows the experimental ground-state energies $E_{gs}$, the excitation energies *E* of the 21$^{st}$ and the 60$^{th}$ state, and the absolute particle separation energies with respect to the liquid-drop ground-state energy [24]. All the experimental data were taken from Ref. [26]. As the most prominent features one observes a strong even-odd effect in the ground-state energies, which almost completely disappears in the excited levels. Thus, the expectation from the super-fluid nucleus model [43, 44] is essentially confirmed. However, a tiny fraction of the even-odd staggering survives in the energies of excited levels. This effect goes beyond the behaviour of quasi-particle excitations depicted in Figure 13 as expected from the super-fluid nucleus model. While blocking effects are expected to destroy the even-odd staggering of the ground-state energies with the first quasi-particle excitations (see Figure 13), the even-odd staggering in the *N=Z* nuclei obviously survives up to excitation energies of several MeV above the ground state. For *N=Z+2* nuclei, this effect seems to be smaller. Unfortunately, the available information from spectroscopy does not reach up to the separation energies. In addition, it is not sure that the experimental information from spectroscopy on the number of levels is complete, in particular at higher excitation energies. The even-odd staggering of the energies of excited levels is an interesting phenomenon by itself, which could have some influence on the observed even-odd staggering in the yields of even-mass nuclei. In addition, there is some even-odd staggering in the absolute particle separation energies which also influences the number of particle-bound states.

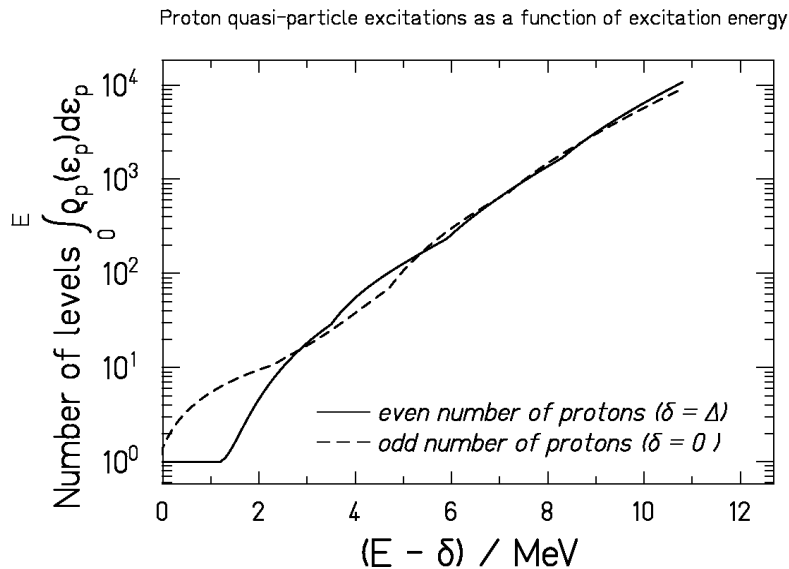

**Figure 13:** Number of proton quasi-particle excitations of two nuclei, $^{100}$Zr and $^{100}$Nb, one with an even and one with an odd number of protons, calculated with Equation (9). Note that the energy scale (*E-δ*) of the even-Z nucleus is shifted by Δ to account for the even-odd effect in the ground-state masses.

These findings give indications that the fine structure in the formation cross sections of light residues, produced in different nuclear reactions, carry information on complex nuclear-structure phenomena. These phenomena might show up in complex properties of the nuclear level density, which go beyond the standard description of the super-fluid nucleus model [44].

## 6. Additional nuclear-structure effects

We have seen that in the case of odd-mass residues, the even-odd staggering in the production cross sections can easily be understood as a manifestation of pairing correlations in the particle separation energies. A simple model, based on the number of particle-bound states in the final products, reproduces the behaviour of nuclei with *N-Z*=odd in all details. However, in order to describe the even-odd staggering in even-mass nuclei, one has to consider not only the available states in the daughter nucleus but also the number of available states in the mother nucleus in the evaporation chain. Only after applying a full evaporation calculation, the behaviour of *N-Z*=even nuclei could be explained. In spite of this success, the particularly strong staggering of the experimental formation cross sections on the *N=Z* chain could not be reproduced.

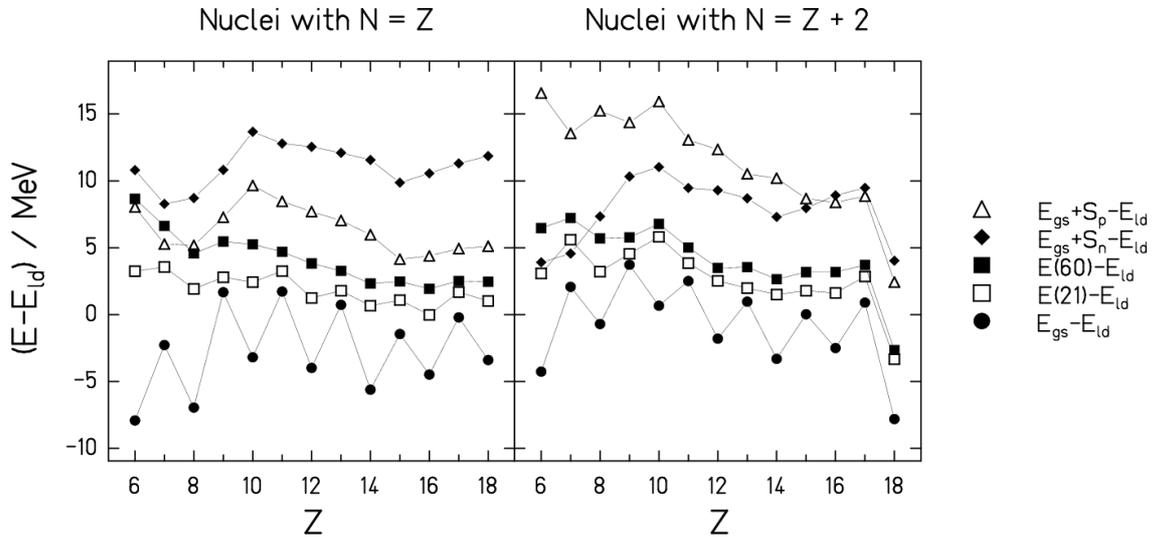

**Figure 14:** The energy of the ground state $E_{gs}$, of the 21$^{st}$ and 60$^{th}$ excited states and the absolute separation energies $E_{gs}+S_\nu$, for *N=Z* and *N=Z+2* nuclei, above the energy of the corresponding liquid-drop [24] ground state $E_{ld}$. The compilation of the data is taken from Ref. [26]. The liquid-drop model does not contain any even-odd effect. The figure demonstrates the gradual decrease of the even-odd structure with increasing excitation energy. The symbols are defined in the figure.

In this Section, which has a rather speculative character, we would like to mention some complex phenomena, which might additionally influence the observed even-odd structures, in particular the enhanced fine structure observed in the *N=Z* nuclei. However, a quantitative analysis is beyond the scope of this article.

### 6.1. Wigner energy and alpha clustering

We have already mentioned that the particularly strong even-odd structure in the production cross sections of *N=Z* nuclei goes in line with an exceptionally strong fluctuation in the binding energies, which is caused by the Wigner energy [39, 40]. As the consequence of the Wigner energy, odd-odd nuclei with *N=Z* are bound by about 30/*A* MeV [41] less than even-even *N=Z* nuclei in addition to the difference of 2Δ expected from pairing. For other nuclei with *N≠Z*, the Wigner energy is a smooth function proportional to |*N-Z*|, and, therefore, has no consequence on even-odd staggering. For a quantitative estimation of the influence of the Wigner energy on the production cross sections of *N=Z* nuclei, however, one needs to describe how the fluctuating Wigner energy in these nuclei behaves as the function of excitation energy.

On the other hand, as even-even *N=Z* nuclei are multiples of alpha particles it is tempting to relate their enhanced production to the alpha clusterisation in nuclei [45, 46]. However, one should be careful with assigning the enhanced even-odd effects in the ground state masses for this class of nuclei to alpha clustering. Jensen *et al.* have analysed experimental ground-state masses using different filters and have not found any trace of alpha clustering [42]. On the other hand, it could be possible that alpha clustering appears for excited states resulting in the enhanced production of even-even *N=Z* nuclei.

### 6.2. Neutron-proton pairing

Also neutron-proton pairing is discussed to play in important role in *N=Z* nuclei [47]. An eventual influence of neutron-proton pairing on the energy of excited levels could be another explanation for the strong even-odd structure in the production yields of the *N=Z* nuclei.

### 6.3. Mean-field contributions to pairing effects

Recently, an interplay between pairing and mean-field effects has been discussed (e.g. [48]). The even-odd mass differences are understood as the sum of the variation of pairing correlations in a given potential, the blocking effect, as discussed above, and the spontaneous breaking of spherical symmetry due to the presence of unpaired particles (Jahn-Teller effect [49]). The second effect is derived to be particularly strong in light nuclei. It is to be expected that the mean-field contribution to the even-odd mass differences also influences excited levels in contrast to the blocking effect, which was schematically considered above. This would explain the general appearance of even-odd differences in the nuclear level densities, (Figure 14), leading to an additional enhancement in the production of nuclei with even proton and neutron number. This effect should decrease with increasing mass of the reaction products.

## 7. Conclusion

Structural effects in the yields of the final products of fragmentation reactions have been investigated. It was shown that the distributions of light fragmentation residues after violent heavy-ion collisions reveal an even-odd staggering of similar magnitude as in the case of low-energy reactions. This was observed in different reactions with different target-projectile combinations and at different beam energies. From a recent experiment with full nuclide identification, this structure has been analysed for cuts along different values of neutron excess *N-Z* and compared with the corresponding structure in the nuclear binding energies. A complex behaviour has been revealed. Odd-mass nuclei with positive *N-Z* values show an enhanced production of odd elements. For *N=Z+5* nuclei, the enhancement is around 40 %. Even-*Z* nuclei along *N=Z* show a particularly strong enhancement in the order of 50% in parallel with a particularly strong even-odd structure in the ground-state binding energies. The production of other even-mass nuclei fluctuates by about 10%: the production of even-even nuclei is slightly enhanced.

A simple statistical model could qualitatively reproduce the staggering of the experimental production yields for light odd-mass residues. This induced us to deduce that structural effects are restored in the end products of hot decaying nuclei, and that the structure is ruled by the available phase space at the end of the evaporation process. On the other hand, the experimentally observed even-odd staggering in the production cross sections of light even-mass nuclei can not be explained with the number of final bound states. It was shown that for *N-Z*=even nuclei, the number of available levels in the mother nucleus also plays a role and is responsible for the staggering in the production cross sections. In the case of heavy products, γ emission becomes a competitive decay channel in the last de-excitation steps and is responsible for the vanishing of the even-odd staggering with increasing mass.

We have also seen that the chain of *N=Z* nuclei appears as a special class of nuclei with increased enhancement in the production of even-even nuclei compared to other chains with *N-Z*=even. Possible origins like the Wigner energy, alpha clustering, and neutron-proton pairing were mentioned.

We conclude that nuclear structure can manifest itself also in the end products of the decay of very hot nuclei. It seems that the fine structure in the production yields from highly excited nuclei is a rich source of information on nuclear-structure phenomena in slightly excited nuclei found at the end of their evaporation process. It is a challenge to quantitatively interpret these results with theoretical models in order to better understand the complex nuclear-structure phenomena behind.

**Acknowledgement**

This work was supported by the European commission in the frames of the EURISOL and the HINDAS projects.